# Large spin-to-charge conversion at room temperature in extended epitaxial Sb$_2$Te$_3$ topological insulator chemically grown on Silicon

Emanuele Longo[1,2,*], Matteo Belli[1], Mario Alia[1], Martino Rimoldi[1], Raimondo Cecchini[1], Massimo Longo[1], Claudia Wiemer[1], Lorenzo Locatelli[1], Gianluca Gubbiotti[3], Marco Fanciulli[2] and Roberto Mantovan[1,**]

[1] CNR-IMM, Unit of Agrate Brianza (MB), Via C. Olivetti 2, 20864, Agrate Brianza (MB), Italy

[2] Università degli studi di Milano-Bicocca, Dipartimento di Scienze dei Materiali, Via R. Cozzi 55, 20126, Milano, Italy

[3] Istituto Officina dei Materiali del CNR (CNR-IOM), Sede Secondaria di Perugia, c/o Dipartimento di Fisica e Geologia, Università di Perugia, I-06123 Perugia, Italy

*emanuele.longo@mdm.imm.cnr.it, **roberto.mantovan@mdm.imm.cnr.it

**Abstract**

Spin-charge interconversion phenomena at the interface between magnetic materials and topological insulators (TIs) are attracting enormous interest in the research effort towards the development of fast and ultra-low power devices for the future information and communication technology. We report a large spin-to-charge conversion efficiency in Au/Co/Au/Sb$_2$Te$_3$/Si(111) heterostructures based on Sb$_2$Te$_3$ TIs grown by metal organic chemical vapor deposition on 4" Si(111) substrates. By conducting room temperature spin pumping ferromagnetic resonance, we measure an inverse Edelstein Effect length $\lambda_{IEE}$ up to 0.75 nm, a record value for 3-dimensional chalcogenide-based TIs heterostructures. Our results open the path toward the use of chemical methods to produce TIs on large area Si substrates and characterized by highly performing spin-charge conversion, thus marking a milestone toward future technology-transfer.

**Introduction**

Information and Communication Technologies (ICT) are deeply changing our lives and working routines, and this trend got remarkably boosted during the Covid-19 pandemic. Governments' digital agendas consider expanding the use of ICT products and services[1] at all levels. In the 2005-2019 period, the number of individuals using the Internet grew from 1.1 billion to 4 billion, representing the 51% of the world population.[2,3] The ever-expanding ICT will have a huge impact in terms of power consumption. In 2020, the electricity consumption due to ICT was ~ 3000 TWh i.e. 11% of the total, with a foreseen increase up to 8000 TWh in 2030.[4,5] This constant increase could have a strong impact on climate change, which is one of greatest challenges of the 21st century.[6] In order to improve the overall efficiency and lower the power consumption of any electronic circuit and device, new materials with enhanced functionalities must be brought to a maturity level.

Topological Insulators (TIs) represent a state of matter in which the material bulk has insulating properties while the surface hosts highly conducting states.[7] In TIs, electrons are characterized by a Dirac-like dispersion energy and very strong spin-orbit coupling determine the electron spin orientation with respect to their momentum thus generating topologically protected surface states (TSS).[7] TIs are therefore considered a very plausible solution to bring spintronics to the next level in the future ICT,[5,8] in which the devices' functionalities can be driven by a collection of spin-orbit coupling phenomena such as spin Hall effects (SHE).[9] Thanks to their TSS, TIs provide an efficient alternative to the typically used heavy metals (HM) for exploiting spin-charge interconversion effects in heterostructures where TIs and magnetic materials are interfaced.[10,11] The second generation of 3-dimensional (3D)-TIs, such as bismuth and antimony chalcogenides-based $Bi_2Se_3$, $Bi_2Te_3$ and $Sb_2Te_3$, is attracting huge interest.[12–14] They are narrow band-gap semiconductors with rhombohedral crystalline structures belonging to the R-3m space group.[12,14] In principle, exploiting TSS in these 3D-TIs requires epitaxial quality thin films, feature most commonly achieved by the widely-reported Molecular Beam Epitaxy (MBE) deposition method,[15–19] with several reports about the use of magnetron sputtering also available.[20–22] In order to fill the gap between research and technology, a firm and decisive effort to develop methods to grow TIs on large-area Si substrates, by simultaneously controlling their functional properties, is highly required. Recently, chemical methods, such as Atomic Layer Deposition, Chemical Vapor Deposition (CVD), and Metal-Organic CVD (MOCVD) were shown to allow cost-effective depositions and complex 3D structures on large areas.[23,24] In a recent review by Zabaveti *et al.*[25] a comparison between growth methods for the synthesis of chalcogenides thin films in terms of their lateral dimension, has showed the clear advantage in using chemical methods (i.e. cost-effectiveness, complex 3D structures).

We recently developed a MOCVD process to grow epitaxial-quality Antimony Telluride ($Sb_2Te_3$) on 4" Si(111) substrates[24] (Supplementary Info. – Fig. S1). When compared to granular-$Sb_2Te_3$ grown on $SiO_2$,[26] the epitaxial-$Sb_2Te_3$ on top of Si(111) shows improved magnetoconductance (MC) performances especially upon proper annealing, providing clearer and more robust TSS (Supplementary Info – Fig. S2). The next fundamental step is therefore to quantify and optimize spin-charge interconversion phenomena at the interface of the developed TIs with magnetic materials.

The use of spin-pumping ferromagnetic resonance (SP-FMR) to investigate spin-to-charge (S2C) conversion at ferromagnets (FM)/HM interfaces has been theoretically described for a long time,[27,28] and widely demonstrated.[29-35] Alternatively, also spin torque – FMR (ST-FMR)[36–38] and second harmonic longitudinal voltage[17,39,40] measurements have been reported. In the case of FM/TIs systems, several reports have recently emerged with studies by ST-FMR,[41–43] spin Seebeck effect,[44] or SP-FMR.[20,29,31,34,35,45,46]

In this work, we report a large S2C conversion occurring at room temperature (RT) in Au/Co/Au/$Sb_2Te_3$/Si(111) heterostructures, by making use of broadband FMR (BFMR), also known as *all-electrical spin wave spectroscopy*, and SP-FMR. In SP-FMR, a pure spin current is generated in the Co layer and perpendicularly pumped into the adjacent 3D-$Sb_2Te_3$, through the Au interlayer, which is found essential for suppressing interfacial non-linear effects due to two magnon scattering (TMS). As a figure of merit for the S2C conversion efficiency quantification, we measure the inverse Edelstein effect length[47] $\lambda_{IEE}$, which is found to range from 0.28 nm to 0.75 nm. These $\lambda_{IEE}$ values are comparable or larger than those previously communicated for FM/TIs structures,[20,29,31,34,35,45] constituting the first report of S2C conversion involving the binary $Sb_2Te_3$. The successful integration of $Sb_2Te_3$ on silicon, opens interesting routes toward the technology transfer of TIs for the future of ICT. Finally, by comparing our study with those obtained so far in FM/TIs systems by SP-FMR, we shed light on the influence of the data-treatment to extract $\lambda_{IEE}$, pointing towards the need for an unified approach to efficiently compare results from different research groups.[17]

## Methodology

Sb$_2$Te$_3$ thin films with a nominal thickness of 30 nm are produced at RT by MOCVD on 4" intrinsic Si(111) wafers (resistivity > 10000 Ω·cm) exploiting an AIXTRON 200/4 system, operating with an ultra-high pure Nitrogen carrier gas and equipped with a cold wall horizontal deposition chamber accommodating a 4" IR-heated graphite susceptor (see Supplementary Info. – Fig.S1). In order to promote an epitaxial order, the Sb$_2$Te$_3$ films are subjected to specific *in-situ* thermal treatments.[24]

The Au(5nm)/Co and Au(5nm)/Co/Au(5nm) capping layers are prepared by e-beam evaporation on pre-cut ~1 x 1 $cm^2$ Sb$_2$Te$_3$ pieces using an Edwards Auto306 facility, producing Au(5nm)/Co($t$)/Sb$_2$Te$_3$ and Au(5nm)/Co($t$)/Au(5nm)/Sb$_2$Te$_3$ heterostructures, with the nominal thickness ($t$) within the 2 - 30 nm range (Fig. 1(a)).

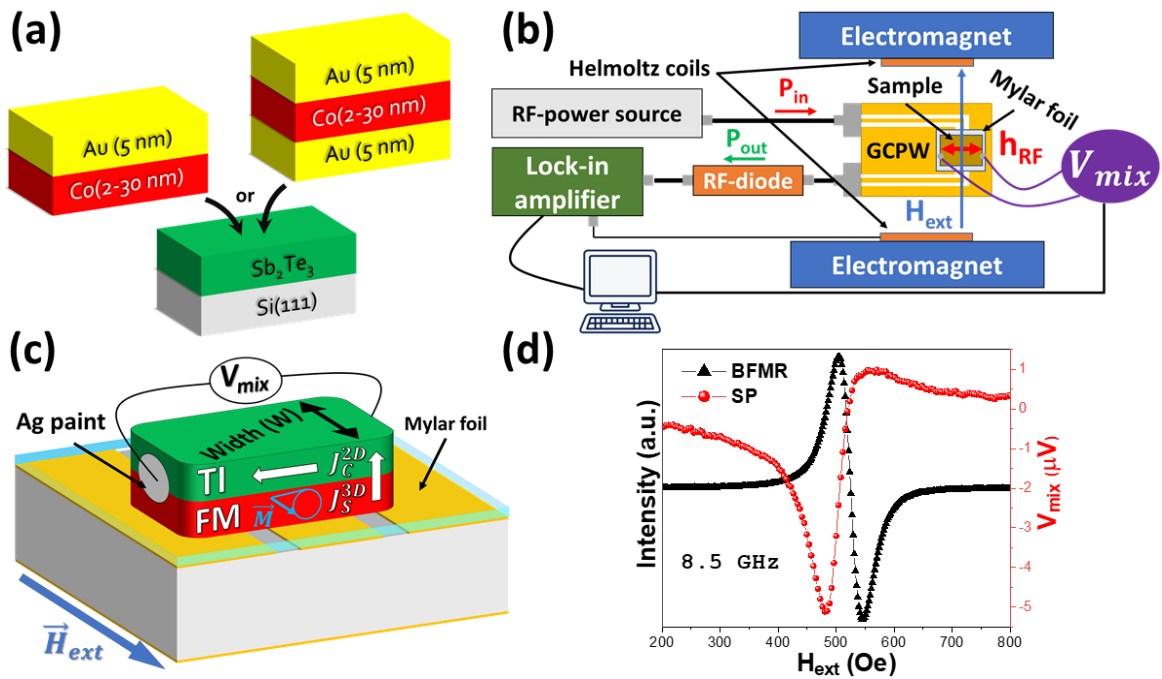

**Figure 1**: (a) Stacking of the investigated heterostructures. (b) BFMR and SP-FMR experimental set up. (c) The geometrical configuration of the sample (width: W) when mounted flip-chip on the GCPW and electrically connected to a nano-voltmeter through Ag wires and Ag paint. The dc-voltage signal ($V_{mix}$) is collected at the sample boundaries. (d) Examples of the acquired FMR (black triangles) and SP-FMR (red dots) signals for Au(5nm)/Co(20nm)/Au(5nm)/Sb$_2$Te$_3$ heterostructures at the fixed RF frequency of 8.5 GHz and fixed RF power of 73 mW. The superposition of the curves demonstrates the direct connection between the two physical phenomena.

The BFMR and SP-FMR experiments are conducted using a home-made setup as depicted in Fig. 1(b), where the sample is positioned between the polar extensions of a Bruker ER-200 electromagnet, maintaining its surface parallel to the external magnetic field ($H_{ext}$) in the so-called "flip-chip" configuration for in-plane (IP) measurements.[48] To induce an oscillating magnetic field in the FM layer, the sample is fixed on a custom grounded coplanar waveguide (GCPW) (Fig. 1(b,c)) connected to a broadband Anritsu RF-source (Supplementary Info. – Fig. S3 and Fig. S4). The FMR signal for a fixed RF frequency is performed by measuring the derivative of the absorption power downstream of the electrical transmission line as a function of $H_{ext}$ through a lock-in amplifier (Fig. 1(b)). In the SP-FMR experimental configuration, the sample edges are connected to a nano-voltmeter with Ag wires soldered with Ag paint and a voltage signal ($V_{mix}$) is measured as a function of $H_{ext}$ (Fig. 1(c)). In Fig. 1(d), an example of the FMR (black triangles) and SP-FMR (red dots) signals detection as recorded in an Au(5nm)/Co(20nm)/Au(5nm)/Sb$_2$Te$_3$ sample at 8.5 GHz, is reported. The two signals are revealed simultaneously and resonate perfectly at the same external magnetic field, demonstrating the correlation between the occurring physical effects.

**Effective spin-mixing conductance in Co/Sb$_2$Te$_3$ heterostructures**

By BFMR, we measure the evolution of the $f_{res}(H_{res})$ curves as a function of the Co thickness for both the Au(5nm)/Co(t)/Sb$_2$Te$_3$ and Au(5nm)/Co(t)/Au(5nm)/Sb$_2$Te$_3$ heterostructures. For each Co thickness, the acquired datasets are fitted to the Kittel formula for the uniform magnetization precession in the IP configuration, as described in the (Supplementary Info. – Fig. S5). Both the sample series show an evolution as a function of the Co thickness in accordance with the Kittel formula, similar to measurements conducted by other groups.[49,50] This underlines the accurate Co thickness control and the overall high magneto-structural quality of the deposited films. In Fig. 2(a) and (b), the linewidth of the FMR signals ($\Delta H$) as a function of the resonant RF frequency ($f_{res}$) is reported for the two series of samples (with and without Au interlayer) and fitted by Eq. (1).

$$\Delta H = \frac{4\pi\alpha}{\gamma} f_{res} + \Delta H_0 \qquad (1)$$

where $\alpha$ represents the damping constant of the FM magnetization, $\gamma$ the gyromagnetic ratio and $\Delta H_0$ the inhomogeneous broadening. The latter parameter provides information about the magneto-structural quality of a FM film, and it is fundamental to confirm the reliability of the physical properties obtained by BFMR.[51] From the best-fit of the of experimental data to Eq. (1), the damping parameter $\alpha$ for each Co thickness in both the Au(5nm)/Co(t)/Sb$_2$Te$_3$ and Au(5nm)/Co(t)/Au(5nm)/Sb$_2$Te$_3$ systems are extracted, and the values are plotted in Fig. 2(c) as a function of the inverse of the Co thickness ($1/t_{Co}$).

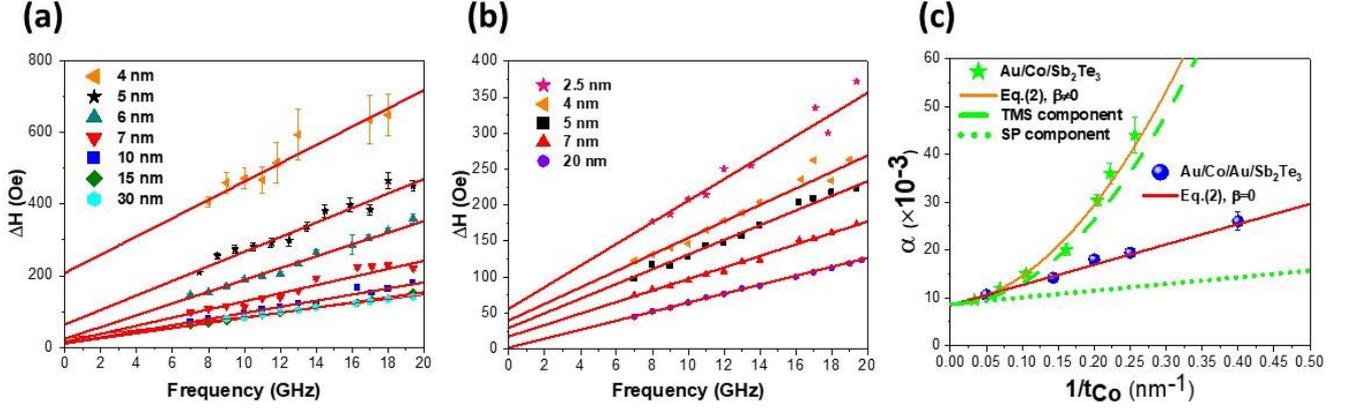

**Figure 2**. $\Delta H(f_{res})$ curves for (a) Au/Co/Sb$_2$Te$_3$ and (b) Au/Co/Au/Sb$_2$Te$_3$ heterostructures, respectively. Error bars in (b) are of the order of the symbols' diameter. (c) Comparison between the $\alpha(1/t_{Co})$ dispersion for the Au/Co/Sb$_2$Te$_3$ (green stars) and Au/Co/Au/Sb$_2$Te$_3$ (blue dots) heterostructures. The orange solid line indicates the fit of the collected data for the Au/Co/Sb$_2$Te$_3$ stack (green stars) using Eq. (2), with $\beta TMS$ as a free parameter. The dashed and dotted green lines represent the TMS and SP components extracted from the orange solid line fit, respectively. The red solid line indicates the fit of the data for the Au/Co/Au/Sb$_2$Te$_3$ structures (blue dots) using Eq.2 where $\beta TMS$ is null.

Typically, in the framework of the SP theory,[28,52] the $\alpha(1/t_{Co})$ curve follows a linear trend as described by the first two terms on the right-hand side of Eq. (2),

$$\alpha = \alpha_{bulk} + Re(g_{eff}^{\uparrow\downarrow}) \frac{g\mu_B}{4\pi M_s t_{FM}} + \beta_{TMS} \frac{1}{t_{FM}^2} \qquad (2)$$

where $\alpha_{bulk}$ represents the damping constant of the bulk material, $\mu_B$ the Bohr magneton, $M_s$ the saturation magnetization, $g$ the g-factor, $t_{FM}$ the thickness of the FM layer and $Re(g_{eff}^{\uparrow\downarrow})$ is the real part of the effective spin-mixing conductance. The latter quantity plays a central role in the description of the SP phenomena, being directly proportional to the spin current density generated in the FM layer and pumped into the adjacent non-magnetic material, here Sb$_2$Te$_3$, at resonance condition.

Clearly, the trend observed for the Au(5 nm)/Co(t)/Sb$_2$Te$_3$ stacks (green data in Fig. 2(c)) does not follow a linear dependence in the whole thickness range. Indeed, by applying the conventional SP fitting model (first two terms in Eq. (2)), an $\alpha_{bulk}= (5 \pm 1) \cdot 10^{-3}$ is obtained, which is in disagreement with the $(8 \div 11) \cdot 10^{-3}$ range expected for bulk Co.[49,53] Being $g_{eff}^{\uparrow\downarrow}$ a fundamental parameter to judge spin pumping functionalities, the observed nonlinearity in the Au(5 nm)/Co(t)/Sb$_2$Te$_3$ system must be carefully addressed in order to avoid the extraction of unphysical $g_{eff}^{\uparrow\downarrow}$ values from BFMR experiments.[53] The non-linear $\alpha$ enhancement can origin from magneto-structural disorder in the Co thin films and/or at the Co/Sb$_2$Te$_3$ interface. Indeed, for the thinnest samples, the obtained inhomogeneous term $\Delta H_0$ shows a slight enhancement when compared to the thicker samples, see Fig. 2(a) and (b). On the other hand, the XRR analysis (Supplementary Info. – Fig. S6) evidences a high chemical-structural quality of the Co layers, suggesting that the divergence observed in Fig. 2(c) for the Au/Co/Sb$_2$Te$_3$ set (green stars) likely has other origins. Actually, L. Zhu *et al.*(2019)[53] has recently reported and analyzed the BFMR response in several FM/Pt heterostructures, pointing out that, in the majority of the studied systems, the SP is a relatively minor contribution to $\alpha$, when measured in the GHz frequency region. Indeed, they suggested that two further terms should be accounted to properly describe the $\alpha(1/t_{Co})$ curve: Spin Memory Loss (SML) and Two-Magnon Scattering (TMS). SML is an interface effect manifesting with an additional linear contribution to that in Eq. (2). Due to SML, the spin current pumped from the precessing magnetization in a FM is partially suppressed at the interface with an adjacent layer, as a result of back-scattering. Recently, the main source of SML was attributed to the presence of an abrupt interruption (i.e. at the interface) between a FM and a material with high SOC, such as HM or TIs.[54] Differently, the TMS is an energy transfer mechanism between the FMR uniform precessional mode and degenerate spin waves.[55–59] As discussed in Refs.[57,60], the source of the TMS is the presence of defects and imperfections at the surfaces and interfaces of FM thin films, which act as a source of scattering for the precessing magnetization. Indeed, the TMS is often related to the morphological and magnetic roughness at the FM/(HM or TIs) interface. According to Ref.[53], the total damping can be seen as $\alpha = \alpha_{bulk} + \alpha_{SP} + \alpha_{TMS}$, thus giving the full expression in Eq. (2), where $\beta_{TMS}$ is the TMS coefficient, proportional to $\left(\frac{K_s}{M_s}\right)^2$ (with $K_s$, $M_s$ as the interfacial magnetic anisotropy density and the saturation magnetization, respectively) and to the density of the magnetic defects at the FM/(HM or TIs) interface.[60] In our system we cannot separate the linear contributions to $g_{eff}^{\uparrow\downarrow}$ coming from SP or SML, and we therefore consider $g_{eff}^{\uparrow\downarrow}$ as totally originated by SP effects. On the other hand, being the linear region of the green dots in Fig. 2(c) negligible when compared to the parabolic TMS terms, we infer a marginal role played by SML to determine our FMR linewidth. From the global fit of the Au/Co/Sb$_2$Te$_3$ data set with Eq.(2), we obtain $\alpha_{Bulk}$ = (8.7± 0.9)· $10^{-3}$, $g_{eff}$ = (0.8 ± 1) · $10^{19}$ $m^{-2}$ and $\beta_{T\,MS}$ = (4.5 ± 0.9) · $10^{-19}$ $m^{-2}$. The $\alpha_{Bulk}$ value perfectly agrees with those expected for bulk Co, thus demonstrating how the inclusion of the TMS contribution is necessary to interpret our FMR data set over the whole range of thicknesses. Therefore, the adopted fitting strategy provides reliable $g_{eff}^{\uparrow\downarrow}$ values, which are comparable to those previously reported in FM/TIs systems (Table 2). In Fig. 2(c), the orange solid line represents the global fit of the Au/Co/Sb$_2$Te$_3$ data set (green stars) with Eq. (2), where the green dashed and dotted lines are the TMS and SP component, respectively. The observation of the SP component (green dotted line) gives an immediate feeling about how this contribution is almost totally hidden by TMS. The presence of TMS in systems made of FM in contact with non-magnetic materials has been previously investigated by means of angular-dependent FMR measurements.[58,61,62] On the other hand, we are not aware of similar reports about the use of BFMR to study the influence of TMS at FM/TIs interfaces, thus showing how TMS must be carefully considered in order to extract $g_{eff}^{\uparrow\downarrow}$ values in SP experiments involving TIs, similarly as in FM/HM heterostructures.[53]

The analysis of the FMR frequency evolution as a function of the applied field by the Kittel formula is reported in the Supplementary Information (Fig. S5(c,d)) for the set of the Au(5nm)/Co(t)/Au(5nm)/Sb$_2$Te$_3$ samples (t= 2.5, 4,5,7,20 nm). The inclusion of the Au interlayer between Co and Sb$_2$Te$_3$ totally suppresses the TMS contribution, blue data in Fig. 2(c), with the $\alpha(1/t_{Co})$ curve now displaying an ideal linear trend. This is directly reflected in lower $\Delta H_0$ values when compared to those extracted for the Au/Co/$Sb_2Te_3$ stack at similar Co thickness (Fig. 2(a) vs (b)). The extracted $\alpha$ values can now genuinely be attributed to SP from Co across the Au(5nm) interlayer into the epitaxial Sb$_2$Te$_3$. Indeed, from the fit of the $\alpha(1/t_{Co})$ data with Eq. (2) (now with $\beta_{TMS}$=0), we obtain $\alpha_{bulk}$ = (8.5 ± 0.2)· $10^{-3}$ and $g_{eff}^{\uparrow\downarrow}$ = (2.1 ± 0.1) · $10^{19}$ $m^{-2}$. The extracted $\alpha_{bulk}$ is in perfect agreement with the expected values[50] thus validating the fitting procedure. The extracted $g_{eff}^{\uparrow\downarrow}$ is well in the $10^{18} \div 10^{20}$ $m^{-2}$ range reported in most of the FM/(HM, TIs) systems probed by SP-FMR (Table 2).

If a FM thin film is in contact with a good spin sink (i.e. HM, TIs), the generation of pure spin currents from FM into HM or TIs, is associated with a high $g_{eff}^{\uparrow\downarrow}$ value. In principle, the insertion of an interlayer between FM and the non-magnetic layer, could lead to a reduction of SP depending on the spin diffusion length ($\lambda_s$) value characterizing the particular interlayer used.[63] On the other hand, in the case of TIs, the direct contact with magnetic materials could also have a detrimental effect on the TSS,[64] which can be otherwise protected with a proper interlayer. Indeed, in FM/HM, TIs systems there are several examples where the presence of chemical intermixing and morphological/magnetic interface roughness has been shown to play a key role in the S2C conversion efficiency.[20,31,32,53,65,66] Therefore, choosing an appropriate interlayer and finding the best trade-off in maintaining the TIs' TSS while keeping an efficient spin transport across the FM/interlayer/TIs interface, is mandatory but also impressively challenging. By comparing our $g_{eff}^{\uparrow\downarrow}$ with other available results (Table 2), it can be concluded that there is certainly still some room to further enhance the spin mixing at the Co/Au/Sb$_2$Te$_3$ interface. A complete overview of different interlayer options to optimize the SP in Co/Sb$_2$Te$_3$-based systems is out of the scope of the present paper and may be the subject of future studies.

**Spin pumping in Au/Co/Au/Sb$_2$Te$_3$ heterostructures**

In a SP experiment a 3D spin current density $J_S^{3D}$ is generated at resonance in the Co layers, longitudinally injected into Sb$_2$T$_3$ across the Au interlayer, and detected through IP SP-FMR.[28,34,52,54,67,68] The general expression for $J_S^{3D}$ (in units of A/$m^2$) is given by Eq. (3).

$$J_S^{3D} = \frac{Re(g_{eff}^{\uparrow\downarrow})\gamma^2 h_{RF}^2 \hbar}{8\pi\alpha^2} \left( \frac{\mu_0 M_S - \sqrt{(\mu_0 M_S)^2 + 4\omega^2}}{(4\pi M_S \gamma)^2 + 4\omega^2} \right) \frac{2e}{\hbar} \qquad (3)$$

where $\hbar$ is the reduced Plank constant, $\omega$ the frequency of the RF-signal, $e$ the charge of the electron and $h_{RF}$ the oscillating magnetic field generated by the GCPW.

Following the spin pumping into the Sb$_2$Te$_3$ layer, a charge current $I_C$ is generated in the Sb$_2$Te$_3$ layer and detected as a potential drop $V_{SP}$ across the measured sample.[69,70] The electronic transport in our Sb$_2$Te$_3$ layers mainly occurs in 2D, as demonstrated by the MC measurements conducted before the Au/Co/(Au) deposition, and interpreted in the framework of the Hikami-Larkin-Nagaoka model (Supplementary Info. – Fig. S2). Therefore, the charge current density $J_C^{2D}$ that is generated by the $J_S^{3D}$ pumping, can be expressed with Eq. (4).

$$J_C^{2D} = \frac{V_{SP}}{WR} \qquad (4)$$

where $W$ is the width of the sample (Fig. 1(c)), $R$ is the sheet resistance as measured separately at four point in the Van der Pauw configuration in the same setup used for MC studies, and $V_{SP}$ is the voltage that is generated across the sample purely due to the SP from Co into the Sb$_2$Te$_3$ layer. The $V_{SP}$ is obtained from the generated transverse $V_{mix}$, being the quantity directly accessible in a SP-FMR experiment (Fig. 1(c)). The first step is therefore to fit the detected $V_{mix}$ with Eq.(5).[71,72]

$$V_{mix} = V_{Sym} \frac{\Delta H^2}{\Delta H^2 + (H - H_{res})^2} + V_{Asym} \frac{\Delta H(H - H_{res})}{\Delta H^2 + (H - H_{res})^2} \quad (5)$$

where $V_{Sym}$ and $V_{Asym}$ are the symmetric and anti-symmetric Lorentzian functions, respectively, $H_{res}$ is the value of the magnetic field at the resonance and $\Delta H$ is the half-width at half-maximum (HWHM). From the SP theory,[27,69,71,73] the symmetric Lorentzian extracted from the fit in Eq.(5) can be originated only from the SP contribution to the $V_{mix}$ curve, and ideally $V_{Symm} = V_{SP}$. However, this term could also contain the thermal Seebeck effect,[44] and in order to extract the pure SP contribution, $V_{SP}$ is typically obtained through Eq. (6).

$$V_{SP} = \frac{V_{Sym}(+H_{ext}) - V_{Sym}(-H_{ext})}{2} \quad (6)$$

The so-called "spin rectification terms" contribute to the $V_{Asym}$ part, being originated from the anisotropic magnetoresistance and anomalous Hall effect in the Au/Co/Au trilayer.[69,71–73] The adopted fitting procedure of the SP-FMR data are reported in the Supporting Information (Fig. S7) for an Au(5nm)/Co(20nm)/Au(5nm)/Sb$_2$Te$_3$ stack.

To assess the intrinsic role played by Sb$_2$Te$_3$ in boosting the S2C conversion efficiency, the new set of samples listed in Table 1 has been specifically synthesized. Indeed, to isolate the contribution purely due to Sb$_2$Te$_3$ to S2C conversion, the growth of the Au/Co/Au FM stack must be conducted on top of both Sb$_2$Te$_3$ and a reference Si(111) substrate simultaneously. This is the only way to quantitatively compare the samples, excluding any potential different aging effect that could take place with and without the Sb$_2$Te$_3$. The choice of 5 nm-thick Co layers is motivated by the need to compare with the current literature reporting on S2C interconversion phenomena in TIs-based systems, where the thickness of the FM layer is typically below 10 nm.[15,17,19,29,35,38,68,74,75] Moreover, a 5 nm-thick Co layer allows clear signals in both BFMR and SP-FMR configurations, which we compare here, to get a comprehensive picture of the S2C conversion occurring in our systems.

**Table 1**. List of samples for the SP-FMR experiments. All the geometrical and electrical quantities used to calculate the S2C conversion efficiency are reported. *W* indicates the width of each sample, *R* the sheet resistance, $V_{SP}$ the effective symmetric Lorentzian extracted from the fits (Eq. 5), and $J_C^{2D}$ the corresponding charge current as calculated with Eq. (4).

| Sample ID | Stack | W (mm) | R (Ohm) | $V_{SP}$ (μV) | $J_C^{2D}$ ($10^{-3}$ A/m) |
|---|---|---|---|---|---|
| S1 | Au(5nm)/Co(5nm)/Au(5 nm)/Sb$_2$Te$_3$/Si(111) | 2.46 ±0.05 | 14 | 6.10 ± 0.07 | 0.178 ± 0.003 |
| S2 | Au(5nm)/Co(5nm)/Au(5 nm)/Si(111) | 2.36±0.05 | 16 | 1.91 ± 0.04 | 0.051 ± 0.002 |
| S3 | Co(5nm)/Au(5 nm)/Si(111) | 2.16±0.05 | 35 | 1.26 ± 0.07 | 0.017 ± 0.002 |

SP-FMR experiments are conducted on all the samples listed in Table 1 by using an RF-power of 132 mW and RF-frequency of 10.5 GHz. Figure 3(a) shows the $V_{mix}$ acquired for sample S1 (red dots), together with the FMR signal for the same sample (black triangles), clearly showing the link between the detected $V_{mix}$ and the FMR response of the system. According to the SP theory[27], by reversing the direction of the applied magnetic field, the DC voltage relative to the SP contribution must change sign. This is observed for all the samples in Table 1, with Fig. 3(b) showing the case of sample S1.

Figure 3(c) summarizes the ($V_{mix}$ - $V_{Offset}$)/RW curves for all the samples in Table 1, and the extracted $J_c^{2D}$ (from Eq. (4)) are depicted in Fig. 3(d) and listed in Table 1. As expected, in our measured $J_c^{2D}$ there is a certain contribution from Au, as demonstrated by the different $J_c^{2D}$ detected in S2 and S3. Nevertheless, the presence of Sb$_2$Te$_3$ in sample S1 provides a gigantic extra contribution to the S2C conversion, with a 250% enhancement when compared to the reference S2 sample.

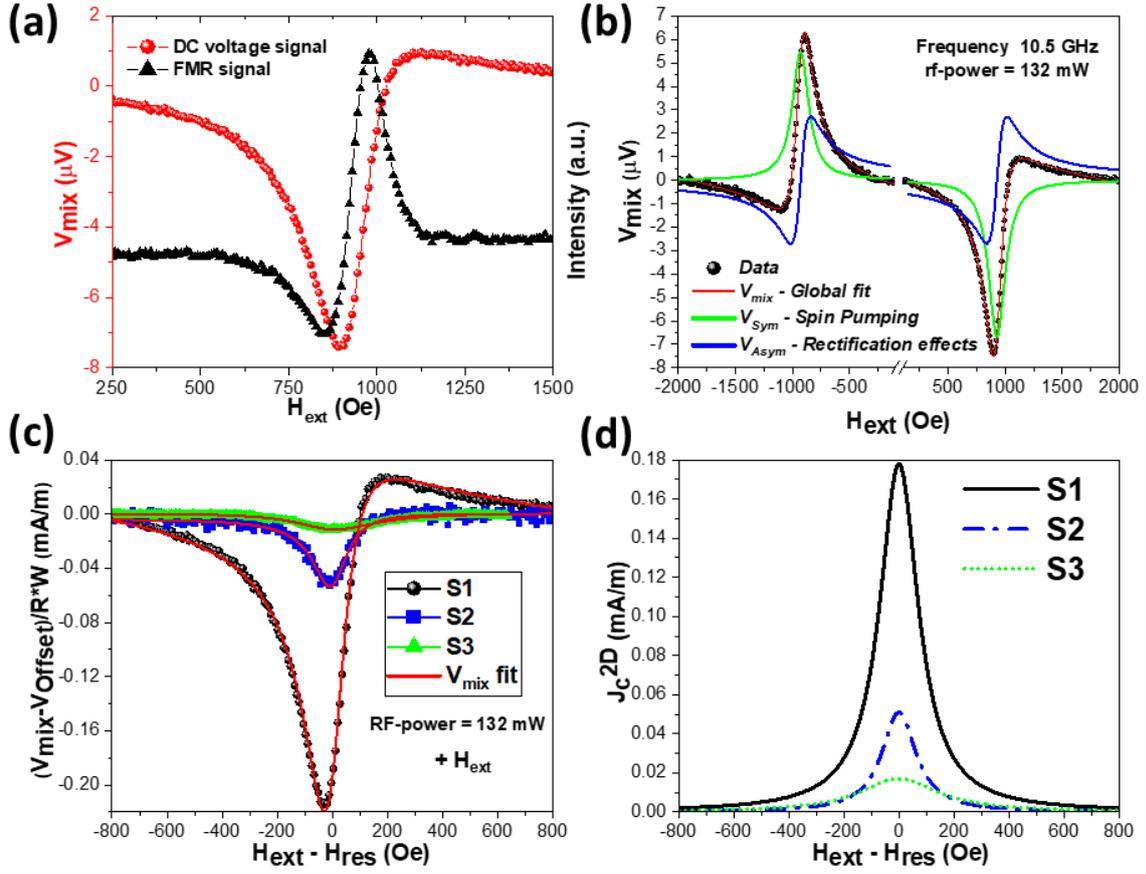

**Figure 3**. (a) SP DC voltage signal for sample S1 acquired at f = 10.5 GHz (red circles). The FMR signal at the same resonance frequency is acquired (black triangles), showing the match between the two signals. (b) The same SP measurement reported in (a) is performed also for negative values of the external magnetic field. Here, it is evident as the asymmetric component $V_{Asym}$ does not depend on the sign of the magnetic field, which is typical for rectification effects due to AMR and AHE. On the other hand, the symmetric component $V_{Sym}$ changes sign upon magnetic field reversal, indicating a magnetic-field dependent spin accumulation. The latter condition is in accordance with SP effects. (c) $V_{mix}-V_{offset}$ signal acquired for samples S1 (black dots), S2 (blue squares) and S3 (green triangles), normalized to the *R* and *W* values for each sample. (d) 2D charge current density ($J_C^{2D}$) extracted from the $V_{Sym}$ component of the $V_{mix}$ signals reported in (a) and calculated using Eq. (4).

The different $J_C^{2D}$ values obtained in samples S2 and S3 indicate that the spin current $J_S^{3D}$ is simultaneously pumped from Co in both the Au layers. Thus, most likely, in sample S1 the spin current pumped into the Au capping layer is reflected at the Au/air interface and then partially absorbed by the $Sb_2Te_3$ substrate. Considering that $\lambda_S$ for Co and Au is ~ 10 nm and ~ 35 nm respectively,[69,76] a tentative sketch of the $J_S^{3D}$ scheme in S1, S2 and S3 is depicted in Fig. 4. Here, the $J_S^{3D}$ backflows at the $Au/Sb_2Te_3$ and Au/Si(111) interfaces, are not considered.

In the case of sample S3, the larger $\Delta H_{S3}$ (175 ± 3 Oe) when compared to both S1 (86.5 ± 0.8 Oe) and S2 (75.5 ± 2.6 Oe), is attributed to the partial Co oxidation due to air exposure. This induces additional structural and magnetic disorder that reflects into a higher magnetic damping.

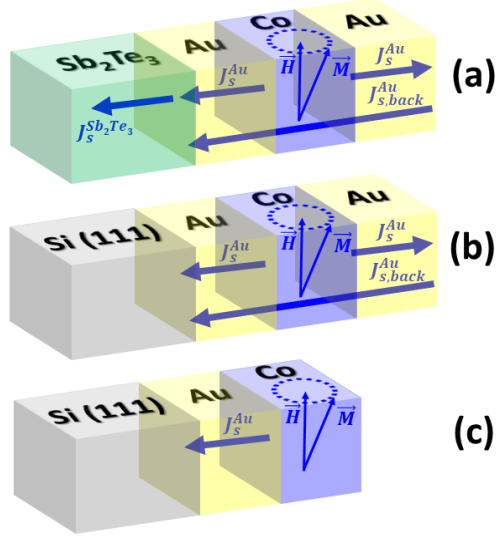

**Figure 4**. Pictorial view of the generated $J_S^{3D}$ current flows in the (a) S1, (b) S2, and (c) S3 samples during the conducted SP-FMR experiments.

**Spin-to-charge conversion efficiency in Au/Co/Au/Sb₂Te₃ stacks**

Our main interest is now to translate the observed additional giant 250% increase in the SP contribution due to Sb₂Te₃ (Fig. 3(c)), into S2C conversion efficiency. In the case of the 2D-type of conduction occurring in our epitaxial Sb₂Te₃ (Supplementary Info. – Fig. S2), the S2C conversion is dominated by the IEE[47], and $\lambda_{IEE} = J_c^{2D} / J_S^{3D}$ is the S2C conversion efficiency figure-of-merit.[29,35,77]

In order to extract the pure contribution due to the Sb₂Te₃, the $g_{eff}^{\uparrow\downarrow}$ required to calculate $J_S^{3D}$ through Eq. (3), must be obtained by considering the additional damping observed in S1 when compared to the reference S2, i.e the additional contribution purely originating from the presence of Sb₂Te₃.[68] This can be done by either considering the difference of the sample damping parameters $\alpha_{S1} - \alpha_{S2}$ (Eq. 7(·))[35,78] or the linewidth of their SP-FMR signals ($\Delta H_{S1} - \Delta H_{S2}$) (Eq. 7(··)).[31,69,77]

$$g^{\uparrow\downarrow}_{eff,Sb2Te3} \dot{=} \frac{4\pi M_s t_{FM}}{g\mu_B}(\alpha_{S1}-\alpha_{S2})$$
$$\ddot{=} \frac{2M_s t_{FM}\gamma}{g\mu_B f}(\Delta H_{S1}-\Delta H_{S2}) \quad (7)$$

In our opinion, the first approach (Eq. 7(·)) is the most accurate since $\alpha$ can be obtained from a linear fit of the FMR broadening change as a function of the resonance frequency, while the second approach (Eq. 7(··)) only considers the difference of the FMR broadening at a fixed frequency. On the other hand, the latter strategy is still at the basis of several reports about SP efficiency in FM/(HM,TIs) systems.[31,70,71,77,79] In fact, the FMR measurements have been typically conducted by adapting cavity electron paramagnetic resonance facilities, with a single RF excitation frequency.[51] It is also not uncommon to see reports of S2C efficiencies extracted from samples having a single FM thickness, and measurements based on a single frequency.[78,79]

In the following, we extract $\lambda_{IEE}$ by following both approaches. Figure 5(a) shows the evolution of the $f_{res}(H_{res})$ curves measured in S1 and S2 and fitted with the Kittel equation for the IP configuration, from which we obtain: $M^{S1}_{eff} = 603 \pm 46 \frac{emu}{cm^3}, g^{S1}=2.64 \pm 0.08$, $M^{S2}_{eff} = 653 \pm 29 \frac{emu}{cm^3}, g^{S1}=2.20 \pm 0.04$.

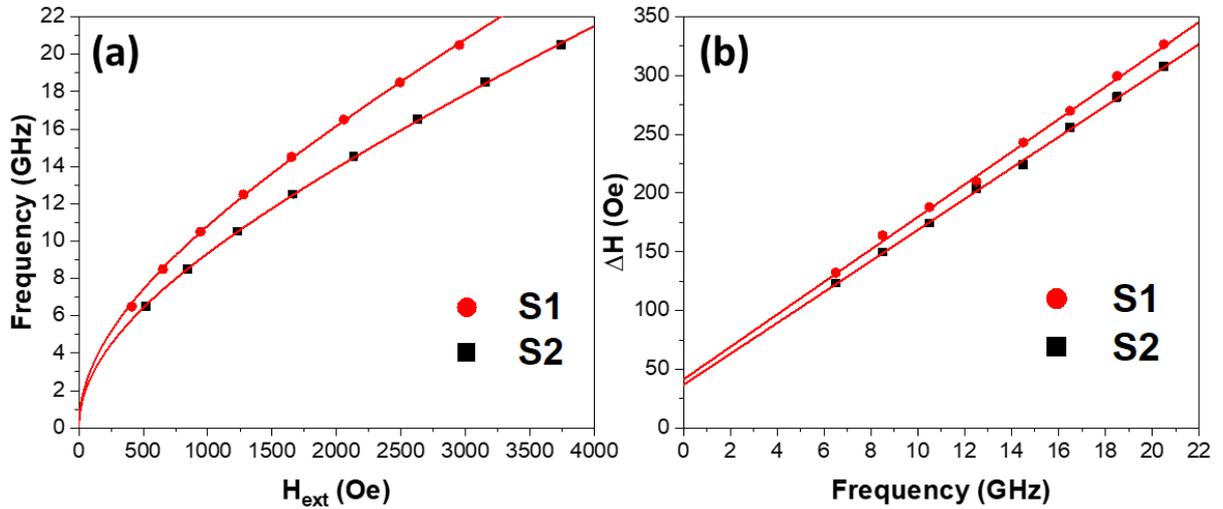

**Figure 5**. BFMR data for samples S1 (red circles) and S2 (black squares). In (a) $f_{res}$ is reported as a function of the resonant magnetic field. From the Kittel fit (red solid line) the $g$-factor and $M_{eff}$ values are extracted for the two samples and reported in the text. (b) The BFMR signal linewidth for samples S1 and S2 is shown as a function of the resonant frequency ($f_{res}$). Here, the damping constants (α) and the inhomogeneous broadening ($\Delta H_0$) are extracted from the linear fit (red solid line) and the values reported in the text.

From the linear best-fit of the FMR signal linewidth as a function of the resonant frequency reported in Fig. 5(b), $\alpha_{S1}$= (25.5 ± 0.6)· $10^{-3}$ and $\alpha_{S2}$= (20.3 ± 0.2)· $10^{-3}$ are extracted. According to Eq. 7(·), these values give $g_{eff,\text{Sb2Te3}}^{\uparrow\downarrow}$ = 8.34 · $10^{18}$ m$^{-2}$, which from Eq. (3) provides $J_S^{3D-Sb2Te3}$ = 6.4 · $10^5$ A m$^{-2}$ as the pure accumulation due to the presence of Sb$_2$Te$_3$ in S1. By considering the $J_C^{2D}$ measured for S1 (Table 1), a $\lambda_{IEE} \sim 0.28\ nm$ value is finally calculated. We now follow the methodology expressed in Eq. 7(··). In particular, we consider the measured $\Delta H_{S1} - \Delta H_{S2} = 11\ Oe$ value at the chosen frequency of 10.5 GHz, which provides a $g_{eff,\text{Sb2Te3}}^{\uparrow\downarrow} = 3.1 \cdot 10^{18}\ m^{-2}$ (Eq.7(··)) and from Eq. (3) $J_S^{3D-Sb2Te3}$ = 2.24 · $10^5$ A $m^{-2}$, finally resulting in $\lambda_{IEE} \sim 0.75\ nm$.

The difference in the obtained $\lambda_{IEE}$ values by following the two approaches of Eq. 7(·) and (··), is relevant. This difference underlines the importance of establishing a common way of reporting S2C conversion efficiencies as measured through SP-FMR. In fact, this represents a necessary step to reliably compare similar FM/(HM, TIs) systems. Moreover, the fitting procedure of the SP-FMR data is not the only controversial aspect that strongly influences the S2C efficiency estimation. As previously pointed out, the FMR signal can be affected by a relevant inhomogeneous broadening contribution, for instance, due to magneto-structural disorder [33] (i.e. magnetic dead layers, presence of different polymorphs in the same FM layer, magnetic roughness) or due to the presence of TMS,[53] revealing that the FMR signal linewidth is not always reflected in an effective SP response. As it follows from Eq. (7), this aspect is directly involved in the calculation of $J_S^{3D}$. Commonly, the subtraction of a proper reference is the only adopted method, and thus considered effective in eliminating all the spurious contributions to the linewidth broadening. Nevertheless, some of us demonstrated that the substrate selection has an important role in governing the magneto-structural properties of a FM thin film, suggesting that unwanted inhomogeneous contribution to the linewidth can be overlooked.[80] Indeed, as recently reported by Nakahashi *et al.*, a more direct and affordable strategy is to measure the $\Delta H(f_{res})$ curve evolution directly from the SP signal.[81]

Table 2 reports a collection of relevant $g_{eff}^{\uparrow\downarrow}$ and $\lambda_{IEE}$ data as obtained by FMR-based methods for heterostructures including TIs, and a selection of HM. The different methods used to interpret the FMR data (Eq. 7(·) vs (··)) are also indicated, with the aim to highlight the need of a standardized procedure of data reporting.

**Table 2.** Summary of $g_{eff}^{\uparrow\downarrow}$ and $\lambda_{IEE}$ values as measured by FMR and SP-FMR (at the indicated temperature T) in stacks with TIs and selected HM. The TIs' and HM's growth methods are also indicated when available. The reported $g_{eff}^{\uparrow\downarrow}$ values are those obtained following the subtraction of corresponding FM's reference samples. The method to extract $\lambda_{IEE}$ following Eq. 7(·) or (··) is also indicated. The data obtained in the present work are reported for comparison.

| Stack | Growth of HM or TI | Thickness (nm) | T (K) | $g_{eff}^{\uparrow\downarrow}$ ($\cdot 10^{19}$ m$^{-2}$) | $\lambda_{IEE}$ (nm) | Analysis by Eq.7 (·) or (··) | REFERENCE |
|---|---|---|---|---|---|---|---|
| Au/Ni$_{80}$Fe$_{20}$ | Not reported | 20/15 | | 0.9 | | (·) | Ref. [82] |
| Pt/Ni$_{80}$Fe$_{20}$ | Not reported | 15/15 | | 3.0 | | (·) | Ref. [82] |
| Pt/Co$_{0.2}$Ni$_{0.8}$ | Sputtering | 6/6 | | ~2 | | (·) | Ref. [83] |
| Pt/Co | Sputtering | | | 3.96 | | (·) | Ref. [83] |
| Pt/Ni$_{0.81}$Fe$_{0.19}$ | Sputtering | 10/10 | RT | 2.31 | | (··) | Ref. [70] |
| Pt/Ni$_{0.81}$Fe$_{0.19}$ | Sputtering | 6/18.5 | RT | 2.4 | | (··) | Ref. [71] |
| (Bi$_{0.22}$Sb$_{0.78}$)$_2$Te$_3$/Ni$_{0.8}$Fe$_{0.2}$ | MBE | 6 QL/12 | | 1.0 | 0.075 | (··) | Ref. [77] |
| Bi$_2$Se$_3$/Ni$_{81}$Fe$_{19}$ | MBE | 20/20 | RT | 1.5 | 0.21 | (·) | Ref. [45] |
| Sn:Bi$_2$Te$_2$Se/Cu/Ni$_{81}$Fe$_{19}$ | Bridgman single crystal synthesis | s.c/5/25 | 40 | - | 0.10÷0.25 | (··) | Ref. [84] |
| α-Sn/Ag/Fe | MBE | 30 ML/2/5 | RT | - | 2.1 | (·) | Ref. [29] |
| Ag/Bi | MBE | 5-20/8 | RT | 1.29÷3.21 | 0.2÷0.33 | (·) | Ref. [34] |
| Bi$_{43}$Se$_{57}$/Co$_{20}$Fe$_{60}$B$_{20}$ | Sputtering | 2 | RT | ~0.7 | 0.32 | (·) | Ref. [35] |
| Bi$_{43}$Se$_{57}$/Co$_{20}$Fe$_{60}$B$_{20}$ | Sputtering | 12 | RT | ~0.7 | 0.10 | (·) | Ref. [35] |
| (Bi,Sb)$_2$Te$_3$/Y$_3$Fe$_5$O$_{12}$ | MBE | 6QL/30 | | | 0.017÷0.035 | (··) | Ref. [46] |
| Tl-Pb/Cu/Ni$_{80}$Fe$_{20}$ | | | 15 | | 0.14 | (··) | Ref. [79] |
| Bi$_2$Se$_3$/Bi/Fe | MBE | 9QL/Bi(n)/13 | RT | ~25÷165.7 | 0.125÷0.28 | (·) | Ref. [85] |
| Bi$_2$Se$_3$/Y$_3$Fe$_5$O$_{12}$ | Sputtering | 4-16/20 | RT | ~0.8-1.36 | 0.11÷0.075 | (·) | Ref. [20] |
| (Bi$_{0.4}$Sb$_{0.6}$)$_2$Te$_3$/Ni$_{0.8}$Fe$_{0.2}$ | MBE | 9/5 | RT | 0.9 | | (·) | Ref. [74] |
| (Bi$_{0.4}$Sb$_{0.6}$)$_2$Te$_3$/Te Ni$_{0.8}$Fe$_{0.2}$ | MBE | 9/4/5 | RT | 2.36 | | (·) | Ref. [74] |
| (Bi$_{0.4}$Sb$_{0.6}$)$_2$Te$_3$/Al/Ni$_{0.8}$Fe$_{0.2}$ | MBE | 9/3/5 | RT | 0.08 | | (·) | Ref. [74] |
| (Bi$_{0.4}$Sb$_{0.6}$)$_2$Te$_3$/Al/Ni$_{0.8}$Fe$_{0.2}$ | MBE | 9/6/5 | RT | 1.71 | | (·) | Ref. [74] |
| (Bi$_{0.4}$Sb$_{0.6}$)$_2$Te$_3$/Ag/Ni$_{0.8}$Fe$_{0.2}$ | MBE | 9/2/5 | RT | 5.71 | | (·) | Ref. [74] |
| (Bi$_{0.4}$Sb$_{0.6}$)$_2$Te$_3$/Ag/Ni$_{0.8}$Fe$_{0.2}$ | MBE | 9/7/5 | RT | 2.83 | | (·) | Ref. [74] |
| Bi$_2$Se$_3$/CoFeB | MBE | 5-10QL/5 | | 1.2-26 | | (·) | Ref. [31] |
| Sb$_2$Te$_3$/Au/Co/Au | MOCVD | 30/5/5/5 | RT | 0.834 | 0.28 | (·) | This work |
| Sb$_2$Te$_3$/Au/Co/Au | MOCVD | 30/5/5/5 | RT | 0.31 | 0.75 | (··) | This work |

If we consider the RT $\lambda_{IEE}$ value extracted with the single-thickness approach, the measured $\lambda_{IEE} \sim 0.75\ nm$ value is, to our knowledge, higher than any other reported S2C conversion efficiency in the chalcogenide-based TIs (i.e. $Bi_2Se_3$, $Bi_2Te_3$ and $Sb_2Te_3$), lower only to that reported for stanine[29] (Table 2). The lower limit $\lambda_{IEE} \sim 0.28\ nm$ is at least of the same order of magnitude (and often higher) of those observed in 3D-TIs produced by MBE or sputtering (Table 2), thus proving the suitability of MOCVD to produce highly performing 3D-TIs on large-area Si substrate. According to the obtained $\lambda_{IEE}$ values, the system here presented may be of interest in the development of magnetoelectric spin–orbit logic devices.[86]

The key to understand the origin for this very large S2C efficiency may lie in the structure and morphology of our $Sb_2Te_3$ layers grown by MOCVD. Recent works have discussed the influence of grain size and grain boundaries in the S2C conversion in $Bi_2Se_3$-based heterostructures as probed by SP-FMR.[20,35,77,87] Interestingly, the granular $Bi_2Se_3$ has been shown to be more efficient in terms of S2C conversion, with a $\lambda_{IEE}$ being 3 times higher than in crystalline $Bi_2Se_3$.[20] As shown by transmission electron microscopy, even though our optimized layers develop an epitaxial nature character, several grain boundaries are still present.[24] In particular, our $Sb_2Te_3$ films are highly ordered and made by compact $Sb_2Te_3$ crystalline grains with an average diameter that can be estimated between 15 and 20 nm (Supplementary Information – Fig. S8). These grain boundaries may locally influence either the $J_S^{3D} \rightarrow J_C^{2D}$ conversion (i.e. SP) and the subsequent longitudinal transport through the TSS. According to Ref.[20], this may be a possible origin for the observed large $\lambda_{IEE}$. This fundamental aspect could be investigated by studying $\lambda_{IEE}$ for different $Sb_2Te_3$ thicknesses, even if the tuning of the $Sb_2Te_3$ thickness through the developed MOCVD process is not straightforward.

Certainly, there is a fundamental role played by the Au interlayer in the observed large $\lambda_{IEE}$. Several groups have already tried to decouple the FM/TIs interface by introducing an interlayer. For instance, in the seminal work of Roja-Sanchez et al. (2016)[29] the introduction of an Ag interlayer in the Au/Fe/Ag/α−Sn structure was proven to be effectively enhance the S2C efficiency by reaching $\lambda_{IEE}$ = 2.1 nm, as extracted from SP measurements at RT. More recently, the thorough study of the S2C conversion efficiency on the spin-orbit torque (SOT) response in different Py/Interlayer/$(Bi,Sb)_2Te_3$ systems have been reported by F. Bonell *et al.*,[74] where the introduction of different metallic spacers (i.e. Te, Ag, Al) has been proven as effective in largely enhancing the S2C efficiency as due to the suppression of the interface intermixing and band-bending. Specifically, by following an accurate chemical-structural description of the Py/Interlayer/$(Bi,Sb)_2Te_3$ interfaces, they have evidenced criticalities concerning the Te out-diffusion from (Bi,Sb)$_2$Te$_3$. In our previous works,[88,89] we reported very similar arguments for what concerns the Te interdiffusion in Fe/Sb$_2$Te$_3$ heterostructures, where a "FeTe" type of bonding at the interface is highly favored. Being FeTe a paramagnetic compound, it could hinder any S2C conversion effect at the interface, or at least largely limit the efficiency of such conversion. As a matter of fact, this is one of the main motivations for our choice of a 5 nm Au buffer layer at the Co/Sb$_2$Te$_3$ interface. The Au interlayer efficiently suppresses several detrimental effects at the Co/Sb$_2$Te$_3$ interface, the main one being certainly the TMS (Fig. 2(c)).

The transport properties of the TSS for several free-standing TIs can be studied by different techniques such as angle-resolved photoemission spectroscopy and scanning tunnel microscopy (STM).[29,90] According to the calculation carried out by Fert and Zhang in Ref.[91], the $\lambda_{IEE}$ can be written as the product $\lambda_{IEE} \equiv v_f \tau_p$, where $v_f$ is the Fermi velocity and $\tau_p$ is the momentum relaxation time, which accounts for the electronic scattering in the TIs bands. The $\lambda_{IEE}$ can be considered as equivalent to the longitudinal mean free path at the metal-TIs interface. The obtained $\lambda_{IEE} \sim 0.75\ nm$ can be regarded as the upper limit for the electronic ballistic transport across the Au/Sb$_2$Te$_3$ interface. This value is lower than those reported for free-standing Sb$_2$Te$_3$ surfaces, where several tens of nanometers have been reported (as in Ref.[90]). In the latter work, a Fermi velocity of $v_f \sim 4.3 \cdot 10^5\ m/s$ has been extracted for the TSS of a crystalline Sb$_2$Te$_3$ thin film, as measured by STM measurements. Assuming $\lambda_{IEE} \sim 0.75\ nm$, we can extract $\tau_p \sim 1.7\ fs$. On the other hand, the presence of the additional Au layer in contact with Sb$_2$Te$_3$, could introduce additional relaxation mechanisms for the Sb$_2$Te$_3$ TSS, which can be at the origin of the discrepancy between the mean free path in the Sb$_2$Te$_3$ layer, when measured through STM or SP-FMR. As suggested in Ref.[29,74], the use of an insulating interlayer in place of a metallic one could solve this problem, preserving more efficiently the TSS and consequently further improving the $\lambda_{IEE}$ value. However, the use of an insulating layer as a spacer between the FM and the TI layers to preserve the TSS still represents an open issue. Indeed, for instance, in Ref.[35] a 2 nm MgO interlayer has been used to demonstrate the suppression of the SP pumping signal, as compared to the same system without interlayer.

**Conclusion**

Room temperature SP-FMR has been successfully employed to measure the S2C conversion occurring in the large-area $Sb_2Te_3$ topological insulator produced by MOCVD on 4" Si(111) wafers. An inverse Edelstein Effect length $\lambda_{IEE}$ from 0.28 nm up to 0.75 nm has been measured. The two values being the outcome of commonly used different methodological analysis. Even the lower observed value is at least comparable (and often larger) than those previously reported in chalcogenide-based 3D-TIs produced by sputtering or MBE. Our results constitute a "year zero" for the use of chemical methods to fabricate TIs for highly efficient spin-charge converters, providing a milestone toward the future realistic technology-transfer. To our knowledge this is also the first report of spin pumping in the binary $Sb_2Te_3$. A further improvement of the observed S2C conversion performances could be achieved by manipulating the Fermi level with appropriate material engineering.[15] Our results also point out the need to standardize the reporting of S2C conversion efficiency as probed by SP-FMR.[92]

**Acknowledgments**

We acknowledge the Horizon 2020 project SKYTOP "Skyrmion-Topological Insulator and Weyl Semimetal Technology" (FETPROACT-2018-01, n. 824123).

**Author Contribution**

E.L. and M.B. developed the BFMR and SP-FMR set-up. E.L conducted all the BFMR, SP-FMR and XRR measurements, and performed the analysis of the FMR data with the assistance of M.B. M.A. conducted the evaporation of all the Au/Co-based heterostructures. M.R. and R.C. conducted the growth of the $Sb_2Te_3$ layers under the supervision of M.L. C.W. contributed to the XRR data analysis. L.L. did the MC and $R_S$ measurements. G.G. conducted BLS on selected samples. M.F. supervised the FMR measurements. R.M. conceived the experiments and coordinated the research activity. E. L. and R.M. wrote the manuscript. All authors discussed the results and reviewed the manuscript and the supplementary information.

# Large spin-to-charge conversion at room temperature in extended epitaxial Sb₂Te₃ topological insulator chemically grown on Silicon

Emanuele Longo[1,2,*], Matteo Belli[1], Mario Alia[1], Martino Rimoldi[1], Raimondo Cecchini[1], Massimo Longo[1], Claudia Wiemer[1], Lorenzo Locatelli[1], Gianluca Gubbiotti[3], Marco Fanciulli[2] and Roberto Mantovan[1**]

[1] CNR-IMM, Unit of Agrate Brianza (MB), Via C. Olivetti 2, 20864, Agrate Brianza (MB), Italy

[2] Università degli studi di Milano-Bicocca, Dipartimento di Scienze dei Materiali, Via R. Cozzi 55, 20126, Milano, Italy

[3] Istituto Officina dei Materiali del CNR (CNR-IOM), Sede Secondaria di Perugia, c/o Dipartimento di Fisica e Geologia, Università di Perugia, I-06123 Perugia, Italy

*emanuele.longo@mdm.imm.cnr.it, **roberto.mantovan@mdm.imm.cnr.it

## *Supplementary Information and Methods*

### Large-area MOCVD-grown epitaxial Sb₂Te₃ thin films

Antimony Telluride (Sb₂Te₃) thin films growth is exploited by MOCVD with an AIXTRON 200/4 system, operating with ultra-high pure Nitrogen carrier gas and equipped with a cold wall horizontal deposition chamber, accommodating a 4'' IR-heated graphite susceptor (*Fig. S1*). Amongst the available antimony and telluride sources, antimony trichloride ($SbCl_3$) and bis(trimethylsilyl)telluride ($Te(SiMe_3)_2$) are selected as MOCVD precursors, because their intrinsic chemical reactivity, unlike precursors such as the most commonly employed $SbEt_3$ and $Te(^iPr)_2$, allows room temperature growth, a significant technical improvement in view of large-scale implementation. Prior to the $Sb_2Te_3$ deposition, the Si(111) substrates are treated with HF(5% in deionized $H_2O$) for 3 min, rinsed with deionized $H_2O$ and dried with $N_2$. Subsequently, the samples are quickly loaded into the atmosphere-controlled glove box of the MOCVD chamber. The depositions are carried out at 25 °C for 90 min at 15 mbar pressure, with a total flow of 5.575 l $min^{-1}$, and setting the precursor vapor pressures at 2.28 and 3.32 $\times 10^{-4}$ mbar, respectively, for $SbCl_3$ and Te$(SiMe_3)_2$.

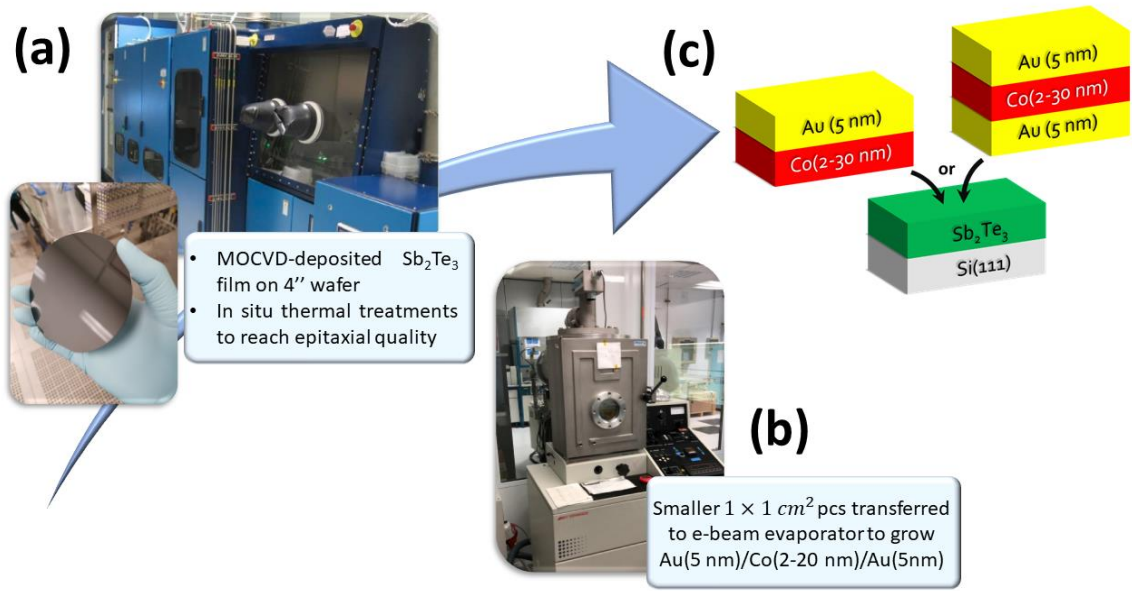

Fig. S1 (a) AIXTRON 200/4 MOCVD system used to grow $Sb_2Te_3$ films on 4'' Si(111) wafers. (b) Edwards Auto360 e-beam evaporator where the Au(5nm/Co(2-30nm)(/Au(5nm)) deposition is done in-situ on top of the epitaxial $Sb_2Te_3$/Si(111) samples; (c) final samples' stack.

In order to obtain the best crystalline quality, the $Sb_2Te_3$ films are subjected to two thermal processes. The first one is carried out prior to the $Sb_2Te_3$ deposition on the Si(111) substrate and performed in situ at 500 °C for 60 min at 20 mbar, with a total $N_2$ flow of 11.000 l $min^{-1}$. The second thermal treatment (post-growth) is performed in situ on the pre-annealed $Sb_2Te_3$/Si(111) structure, according to the following recipe: (1) heating ramp: 5.575 l $min^{-1}$ $N_2$ flow, 900 mbar, from RT to 300 °C in 10 min; (2) annealing: 5.575 l $min^{-1}$ $N_2$ flow, 900 mbar, 300 °C, 15 min; (3) cooling ramp: 1.500 l $min^{-1}$ $N_2$ flow, 990 mbar, from 300 °C to 200 °C in 20 min, from 200 °C to 100 °C in 35 min, from 100 °C to 50 °C in 20 min. As a result, 30 nm thick highly crystalline $Sb_2Te_3$ thin films are obtained. Following the MOCVD of Sb$_2$Te$_3$, substrates are cut into ~ 1 x 1 cm$^2$ pieces and immediately transferred to the Edwards Auto306 e-beam evaporation facility, where the Au/Co and Au/Co/Au bi- and trilayers are deposited all in situ. In all the processes the starting value of the vacuum in the deposition chamber is in the range of $5 \cdot 10^{-7} - 10^{-6}$ Pa. For each evaporated element, the electronic gun deposition current and the value of the vacuum in the growth chamber during the process are: Au 120 mA - $7.8 \cdot 10^{-6}$ Pa; Fe 80 mA - $4.6 \cdot 10^{-6}$ Pa and Co 55 mA - $4.6 \cdot 10^{-6}$ Pa.

# Magnetotransport measurements on Sb$_2$Te$_3$ thin films

Magnetoconductance (MC) measurements constitute a powerful tool for the investigation of the topological properties of a TI. The typical MC curve has a parabolic shape, but in specific materials, due to quantum effects dominating at low magnetic field, a deviation from the canonical parabola can be observed. In particular, the latter phenomenon has been described by Hikami, Larkin and Nagaoka in 1980 with the formulation of the "HLN model",[1] where the electronic weak antilocalization (WAL) and weak localization (WL) effects are linked with the possible existence of topological conductive channels. Mathematically, the HLN theory for our systems can be summarized with the following equation:

$$\Delta G_s = -\alpha \frac{e^2}{\pi h}\left[\psi\left(\frac{1}{2}+\frac{B_\phi}{B}\right) - \ln\left(\frac{B_\phi}{B}\right)\right]$$

where ψ is the digamma function, $e$ the electronic charge, $\boldsymbol{B_\Phi} = \frac{\hbar}{4el_\phi}$ is the dephasing field, $\boldsymbol{\ell_\Phi}$ the spin coherence length of the electron and α is connected to the number of conducting topological channels. In Fig. S2 the MC signal as a function of the external magnetic field applied perpendicular to the sample surface is reported for the MOCVD-grown Sb$_2$Te$_3$ thin films at 5 K. From the fit of the data with the HLN equation, α= -0.25 and l$_\phi$ = 58 nm values are extracted and reported in the inset of Fig.S2. A negative α is associated with the presence of WAL arising from the presence of strong spin-orbit coupling (SOC) and low magnetic scattering.[2] WAL is thus well accepted as a proof of the existence of topological surface states (TSS), that inherently show high SOC and high mobility. In a thin film, the α predicted for the conduction arising solely

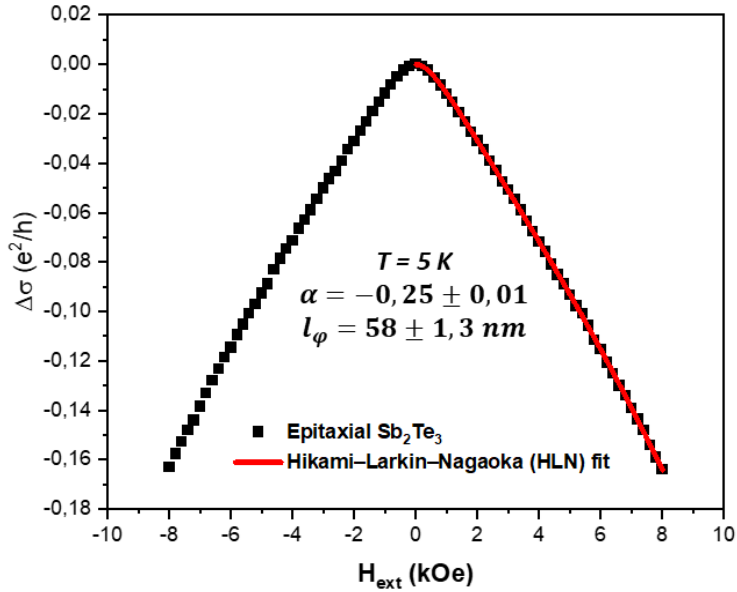

*Figure S2: MC curve for an epitaxial Sb$_2$Te$_3$ thin film deposited by MOCVD and acquired with the magnetic field perpendicular to the sample surface. Here Δσ represents the MC referred to its value at zero field and expressed in unit of e$^2$/h. The red solid line represents the fit of the acquired data (black squares) with the HLN curve. Being the MC curve symmetric with respect to the y-axis, in order to increase the precision of the fit, the latter is performed only for the positive values of the external magnetic field and the fitted data are*

from TSS is -1 if both the interfaces participate to transport and -0.5 if just one of the two surfaces is involved. Chalcogenide based Tis, such as $Sb_2Te_3$, are also heavy materials, where the SOC is relevant, and the bulk states are relatively conductive. For this reason, separating the bulk and surface contribution to the WAL is challenging. To clarify the origin of WAL the magnetic field could be applied also in the plane of the sample, because in this configuration any MC contribution is attributed to bulk states.[3] Our results (not shown here) for α indicate that there exists a mixed contribution of WAL and WL and the measurement performed with the field applied in the film plane suggests that bulk states are not contributing to WAL, but just to WL. In this scenario, α = -0.25 is attributed to a combination of WAL, arising from the TSS, which would give a value of -0.5 and WL, arising from the bulk state, which tends to increase α.

If compared with the granular $Sb_2Te_3$ thin films studied by Cecchini *et al.* [4], where a value of α = -0.01 at T= 5 K has been reported, the topological properties of the epitaxial $Sb_2Te_3$ thin films investigated in this manuscript are largely enhanced, demonstrating the effectiveness of the performed thermal treatments.[5] As discussed in the main text, despite the encouraging results already obtained in terms of spin to charge conversion, it could be possible to further suppress the bulk conductive states contribution, tuning the position of the Fermi level by doping as reported in Ref.[6].

## Home-made grounded coplanar waveguide

The BFMR measurements are performed using a home-made facility obtained from the customization of a Bruker ER-200 instrument, originally adopted for Electron Paramagnetic Resonance (EPR) measurements. The setup is composed by a broadband Anritsu-MG3694C power source (1-40 GHz), which is connected to a home-made grounded coplanar waveguide (GCPW), where the ferromagnetic sample is mounted in a flip-chip configuration (with the FM film close to the CPWG surface), with a 20$\mu m$ thick mylar foil placed in between, to avoid the shortening of the conduction line. The GCPW is connected to a rectifying diode (Wiltron, Model 70KB50 (NEG), 1 - 26.5 GHz, 20 dbm MAX) which converts the RF-signal into a continuous DC-current, in turn sent to a lock-in amplifier for the signal detection. The GCPW is the RF-component used to generate the oscillating $h_{RF}$ magnetic field. A GCPW consists essentially of a central conductor of width $w_s$ which carries the RF-current (signal line, S) and two ground planes (G) separated from the signal line by an air gap of thickness $w_{sg}$ (**Fig. S3.**).

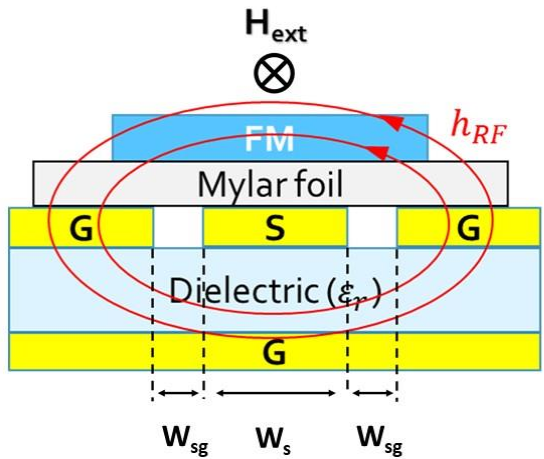
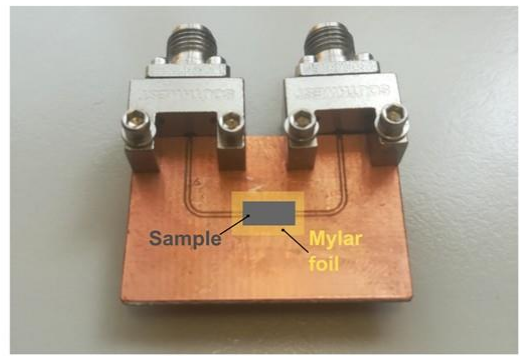

*Figure S3: (a)Illustration of a FM sample positioned in the flip-chip orientation on a GCPW. S and G indicate the signal and the ground conduction lines, respectively.* $h_{RF}$ *represents the oscillating magnetic field produced by the RF-current. (b) Image of the home-made U-shaped GCPW, where the positions of the sample and of the Mylar foil used during the measurements are highlighted. The Mylar foil is located between the sample and the board, to avoid the electrical shortening of the line. To connect the GCPW with the RF power source, two Southwest connectors for high frequency application are used.*

In order to extract the $h_{RF}$ value produced by the GCPW for a fixed RF power, we model the GCPW geometry and calculate the $h_{RF}(z)$ function, where z is the height from the GCPW surface. In Fig. S4 a sketch of the model we adopt is reported together with the canonical Ampere's Law used for the calculation. For a precise evaluation of the position of the sample along the z-axis, measurements of the effective thickness of the GCPW and of the mylar foil are conducted. The SP results reported in the main text are obtained for a fixed RF power of 132 mW, which originates at the sample an oscillating magnetic field of $h_{RF}$ = 0.95 Oe.

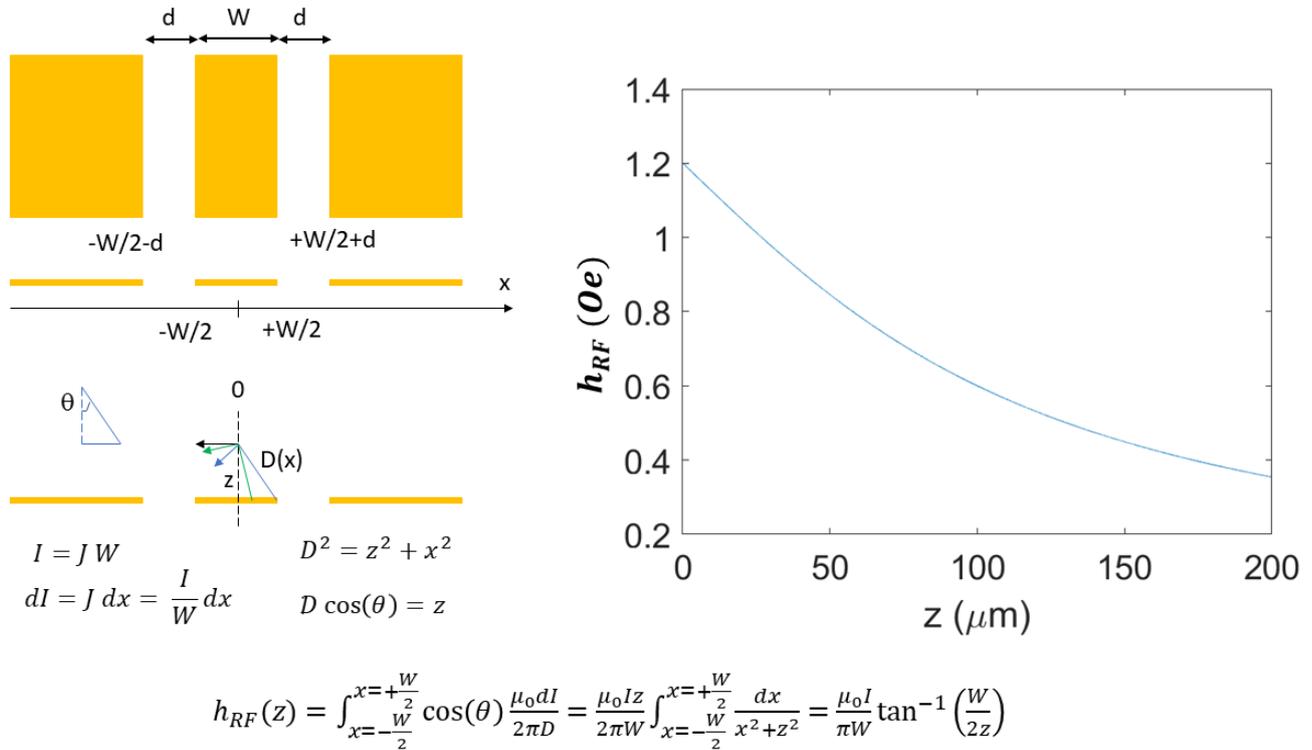

$$h_{RF}(z) = \int_{x=-\frac{W}{2}}^{x=+\frac{W}{2}} \cos(\theta) \frac{\mu_0 dI}{2\pi D} = \frac{\mu_0 I z}{2\pi W} \int_{x=-\frac{W}{2}}^{x=+\frac{W}{2}} \frac{dx}{x^2+z^2} = \frac{\mu_0 I}{\pi W} \tan^{-1}\left(\frac{W}{2z}\right)$$

*Figure S4: Calculation of the oscillating magnetic field $h_{RF}$ produced with the GCPW for a 73 mW RF power.*

## BFMR measurements on the Au(5nm)/Co(t)/$Sb_2Te_3$ and Au(5nm)/Co(t)/Au(5nm)/$Sb_2Te_3$ samples

In Fig. S5 (a,b) the evolution of the $f_{res}(H_{res})$ plots as a function of the Co thickness for the Au(5 nm)/Co(t)/$Sb_2Te_3$ stacks is shown. Here, for each Co thickness the acquired dataset (colored squares) is fitted with the Kittel equation $f_{res} = \frac{\gamma}{2\pi}\sqrt{H_{res}(H_{res} + 4\pi M_{eff})}$ (red solid lines) and the $M_{eff}$ and g-factor values extracted. Due to the weak FMR signals for the sample with Co(2 nm) (pink squares in Fig.3(a)), in this case, the $f_{res}(H_{res})$ signal is acquired by Brillouin Light Scattering measurements, as discussed in the section below. In Fig.S5 (b) the values of $M_{eff}$ extracted by the fit showed in Fig. S5(a) are plotted as a function of the inverse of the Co thickness. Here, from the fit of the data (black squares) using the equation $4\pi M_{eff} = 4\pi M_{eff} - \frac{2K_S}{M_s t_{Co}}$ (red solid line), we obtain $M_S$ = 1030 ± 70 $Oe$ and $K_S$ = 0.99±0.29 $erg/cm^2$.

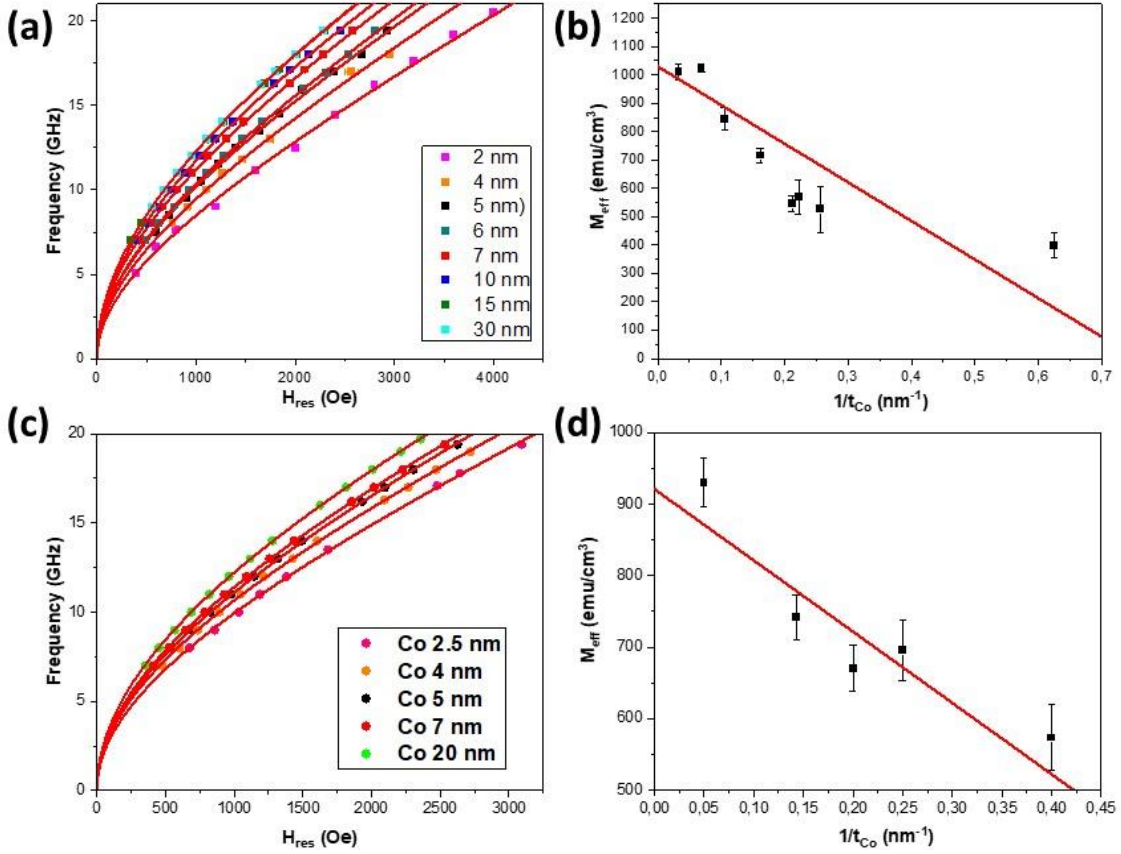

**Figure S5:** (a) Dependence of the $f_{res}(H_{res})$ signal as a function of the Co thickness (colored squares) in the Au(5nm)/Co(t)/$Sb_2Te_3$ samples. The solid red lines indicate the data fit with the Kittel equation for the IP geometry. The data for the 2 nm thick Co samples are acquired by BLS measurements. (b) Values of $M_{eff}$ extracted by the fits in (a) and plotted as a function of the inverse of the Co thickness (1/$t_{Co}$)) measured by XRR. The extracted values are $M_s$ = 1030 ± 70 $Oe$ and $K_s$ = 0.99 ± 0.29 $erg/cm^2$. (c) $f_{res}(H_{res})$ signal for each Co thickness (colored circles) for the Au(5nm)/Co(t)/Au/$Sb_2Te_3$ stacks. (d) Values of $M_{eff}$ extracted by the fits in (c) and plotted as a function of the inverse of the nominal Co thickness (1/$t_{Co}$)). From the analysis of the Kittel curves $M_S$ = 921 ± 55 $emu/cm^3$ and $K_S$ = 0.58±0.18 $erg/cm^2$ are extracted.

For the Au/Co/$Sb_2Te_3$ stack, the value of the extracted Co g-factors varies with the Co thickness, but not with a clear trend. As pointed out in Ref. [7], such variation can be attributed to both the difficulty to extract this value from an IP BFMR configuration, due to the non-linear dispersion of the $f_{res}(H_{res})$ curve and to the possible modification of the properties of the Co interfaces. Nevertheless, the g-factor values are in the range of $g^{Co}$ = 2.37–2.64, which are typical for Co thin films.[8] The $M_{eff}$ value for each sample and further relevant parameters are summarized in Table S5 below, along with the values of the nominal and real thicknesses of the Co thin films (measured by X-ray Reflectivity), the g-factor, the inhomogeneous broadening $\Delta H_0$ and the damping constant $\alpha$, as extracted from Fig.2 in the main text.

*Table S5: Summary of the main quantity extracted from the fits of the BFMR data reported in Fig. S5 and Fig.2 in the main text, and a comparison with real thicknesses extracted by XRR experiments.*

| Nom. Thick. (nm) | Real. Thick. $t_{Co}$ (nm) | $t_{Co}^{-1}$ $(nm)^{-1}$ | g-factor | $M_{eff}$ (emu/$cm^3$) | $\Delta H_0$ (Oe) | $\alpha$ ($10^{-3}$) |
|---|---|---|---|---|---|---|
| 2 | 1.6 | 0.625 | 2.45 ± 0.09 | 399 ± 44 | | |
| 4 | 3.9 | 0.256 | 2.45 ± 0.14 | 526 ± 82 | 207 ± 19 | 44 ± 3 |
| 5 | 4.5 | 0.222 | 2.54 ± 0.1 | 569 ± 61 | 64 ± 13 | 36 ± 2 |
| 6 | 4.7 | 0.211 | 2.64 ± 0.05 | 545 ± 28 | 24 ± 7 | 30 ± 1 |
| 7 | 6.2 | 0.161 | 2.53 ± 0.04 | 716 ± 27 | 15 ± 7 | 20 ± 1 |
| 10 | 9.5 | 0.105 | 2.46 ± 0.05 | 845 ± 41 | 11 ± 4 | 15 ± 0.6 |
| 15 | 14.5 | 0.069 | 2.34 ± 0.01 | 1024 ± 16 | 11 ± 2 | 12 ± 0.2 |
| 30 | 30 | 0.033 | 2.37 ± 0.03 | 1010 ± 28 | 25 ± 2 | 9.5 ± 0.3 |

In Fig. S5(c,d) a complete BFMR study on the Au(5nm)/Co(t)/Au(5nm)/Sb$_2$Te$_3$ stacks is reported. The evolution of the data shows the high quality of the whole set, being in accordance with the FMR theory. Moreover, as discussed in the main text, the parameters extracted from these measurements demonstrate the high magnetic quality and the thickness control of the investigated samples. From the analysis of the Kittel curve the $M_S$ = 921 ± 55 $emu/cm^3$ and $K_S$ = 0.58±0.18 $erg/cm^2$ values are extracted. These values are lower than those extracted for the Co samples directly in contact with the $Sb_2Te_3$ layer. A possible reason can be attributed to the fcc crystalline structure of the Au substrate, which could promote the formation of a higher fraction of cubic crystalline grains in the polycrystalline film, as compared to the same Co deposition on top of the hexagonal $Sb_2Te_3$, which typically develops a hexagonal-phase.[9,10] Indeed, as also reported in Ref.[8], for bulk fcc-Co, $M_S \sim$ 1100 $emu/cm^3$, which is lower than in the hex-Co ($M_S \sim$ 1400 $emu/cm^3$). On the other hand, the $K_S$ values are in accordance with previous studies on Au/Co/Au sandwiches,[11] suggesting that the Co magnetic moment remains close to the bulk value also for very thin Co thicknesses (down to 2.5 nm in this study). A confirmation of the homogeneity of the Co electronic structure over the whole range of thicknesses values is given by the poorly dispersed values for the g-factors, which are all close to $g \sim$ 2.5 (Table S6), compatible with typical values for Co thin films.[8,12]

In Table S6 the parameters extracted from Fig. S5 (c,d) and Fig.2 in the main text are reported for each sample, besides the nominal thickness of the Co layer.

Table S6: Summary of the main quantities relative to the Au(5nm)/Co(t)/Au(5nm)/Sb$_2$Te$_3$ stacks extracted from the fits in Fig.S5(c,d) and Fig. 2 in the main text.

| Nom. Thick. (nm) | $t_{Co}^{-1}$ $(nm)^{-1}$ | g-factor | $M_{eff}$ (emu/$cm^3$) | $\Delta H_0$ (Oe) | α ($10^{-3}$) |
|---|---|---|---|---|---|
| 2.5 | 0.4 | 2.48 ± 0.09 | 573 ± 46 | 67 ± 14 | 26 ± 2 |
| 4 | 0.25 | 2.43 ± 0.05 | 695 ± 42 | 39 ± 6 | 19 ± 0.9 |
| 5 | 0.2 | 2.58 ± 0.05 | 670 ± 32 | 30 ± 5 | 18 ± 0.8 |
| 7 | 0.14 | 2.54 ± 0.05 | 742 ± 31 | 17 ± 1.9 | 14 ± 0.3 |
| 20 | 0.05 | 2.45 ± 0.04 | 930 ± 34 | 2 ± 0.8 | 11 ± 0.2 |

**Brillouin Light Scattering (BLS) measurements general details**

The BLS is an optic technique that makes possible the detection of spin waves traveling within a ferromagnetic film. It is based on the inelastic scattering of monochromatic light from thermally excited spin waves where both energy and momentum are conserved.[14] In order to be used as a complementary analysis to the BFMR, the BLS experiments were conducted by focusing the laser beam at normal incidence upon the sample while the external magnetic field (H$_{res}$), applied parallel to the sample surface, was swept from 3500 to zero Oe (as shown in Figure S5). In such a configuration, BLS is totally equivalent to the BFMR adopted in this work.[13,14] Thanks to the high sensitivity of BLS technique, we were able to detect spin waves in ultra-thin ferromagnetic thin films (less than 2 nm)[15], often too weak to be easily detected by a BFMR measurement.

**X-Ray Reflectivity (XRR) measurements and summary of the main BFMR parameters for Au(5nm)/Co(t)/Sb2Te3 stacks**

In Fig. S6 (left) the XRR measurements for the Au(5nm)/Co(t)/Sb2Te3 stacks are reported. As it can be observed, the XRR model fits almost perfectly the collected data for all the Co thicknesses, witnessing the reliability of the ferromagnet deposition process. For each layer composing the structure, the thickness, electronic density and roughness are summarized in the table reported in the left side of Fig. S6.

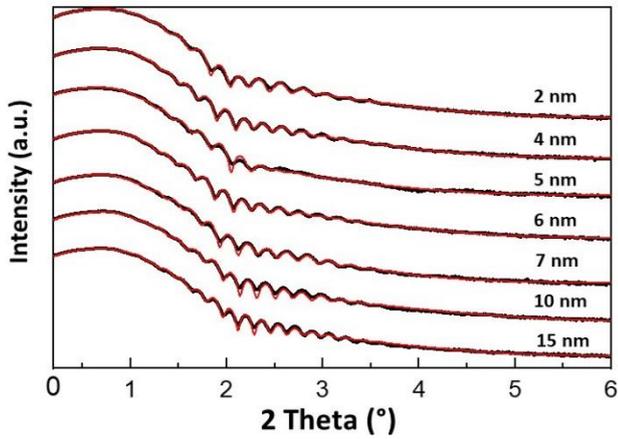

| | Nominal Thickness (nm) | Measured Thickness (nm) | $\rho$ ($e^-/\text{Å}^3$) | Roughness (nm) |
|---|---|---|---|---|
| Sb$_2$Te$_3$ | 30 | 31 ± 3 | 1.6 ± 0.1 | 1.35 ± 0.3 |
| Au | 5 | 4.5 ± 0.5 | 4.2 ± 0.1 | 1.5 ± 0.1 |
| Co | 2 | 1.6 | 2.15 | 1.4 |
| | 4 | 3.3 | 2.15 | 1.57 |
| | 5 | 4.1 | 2.15 | 1.6 |
| | 6 | 4.73 | 2.15 | 1.77 |
| | 7 | 6.2 | 2.15 | 1.52 |
| | 10 | 9.5 | 2.15 | 1.5 |
| | 15 | 14.1 | 2.15 | 1.57 |

*Figure S6: (a) XRR collected data for the Au(5nm)/Co(t)/Sb$_2$Te$_3$ stacks (black lines). The red solid lines indicate the fit of the XRR data with the best model for a multilayered structure. (b) Summary of the thickness, roughness and electronic density for each layer.*

## SP-FMR fitting procedure and power dependence of the spin pumping signal

In order to test the reliability of the experimental setup and the fitting strategy, SP experiments are recorded on various samples with different Co thicknesses. As an example of the adopted fitting procedure, in Fig. S8(a) the SP signal for an Au(5nm)/Co(20nm)/Au(5nm)/Sb$_2$Te$_3$ stack. According to the SP theory[16], the symmetric and anti-symmetric components of the SP signal should be linearly dispersed as a function of the RF power. As an example, in Fig. S8(b) the acquired $V_{mix}$ curves for sample S1 are reported for the fixed frequency of 10.5 GHz as a function of the RF power. In Fig.S8(c), the $V_{Sym}$ and $V_{Asym}$ values extracted from panel (b) are plotted as a function of the RF power, showing their linear behavior. The SP arises as a consequence of the magnetization-precession relaxation, which generates a pure spin current in the FM layer proportional to the

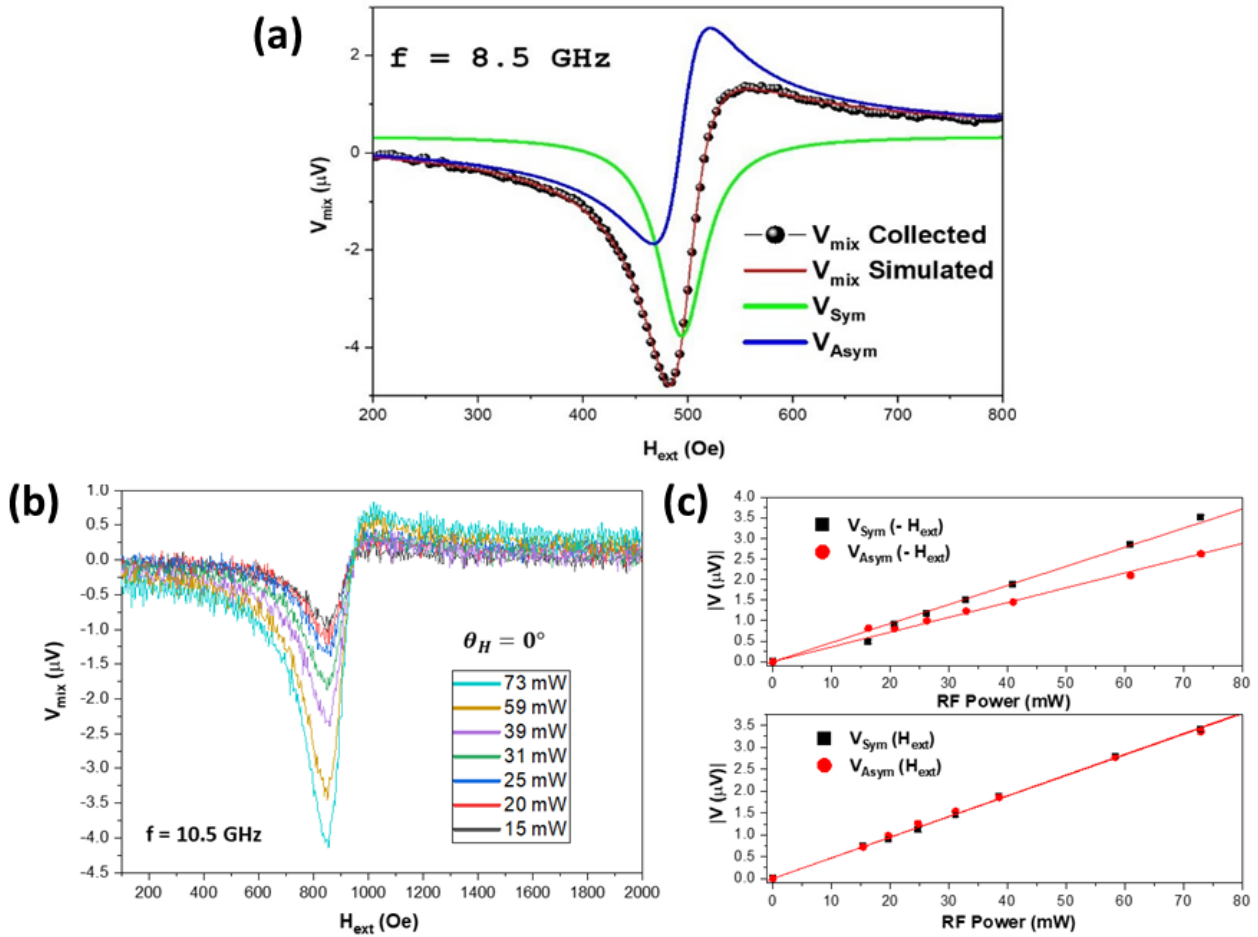

FigureS7: SP-FMR signal for a Au/Co (20 nm)/Sb$_2$Te$_3$ sample. The black circles represent the acquired voltage signal under resonance ($V_{mix}$). The red solid line is the global fit with Eq. (5) in the main text, which is decoupled in the symmetric ($V_{Sym}$) and anti-symmetric ($V_{Asym}$) components indicated by the green and blue solid line, respectively. (b) and (c) show the $V_{mix}$ signal and the $V_{Asym}$ /$V_{Sym}$ components exas a function of the RF power difference of the damping terms, as a result of Eq.3 and Eq.6 in the main text. The DC component of $J_S^{3D}$ is proportional to the projection of the $\vec{M} \times \frac{d\vec{M}}{dt}$ term on the external magnetic field direction. According to the SP theory,[17] such projection is proportional to the square of the magnetization precession amplitude. Thus, $J_S^{3D}$, as well as the IEE DC voltage signal, is proportional to the square of the magnetization precession amplitude or to the square of the applied microwave amplitude. In virtue of that, the IEE signal $V_{Sym}$, should be linear in the RF power, which is consistent with the trend observed in Fig.S8.

## Atomic Force Microscopy on epitaxial Sb$_2$Te$_3$ substrate

Fig. S9 displays the AFM acquisition of the epitaxial Sb$_2$Te$_3$ surface. The surface is characterized by small crystallites with a diameter of about $15-20\, nm$ organized in bigger agglomerates. Due to the finite size of the AFM scanning tip, the estimation of the smaller grain diameter must not be considered as a fully quantitative measurement. The images in Fig. S9 aim to provide a qualitative, but also more complete description of the Sb$_2$Te$_3$ surface morphological properties, which are likely related to the efficient spin-to-charge conversion discussed in the main text.

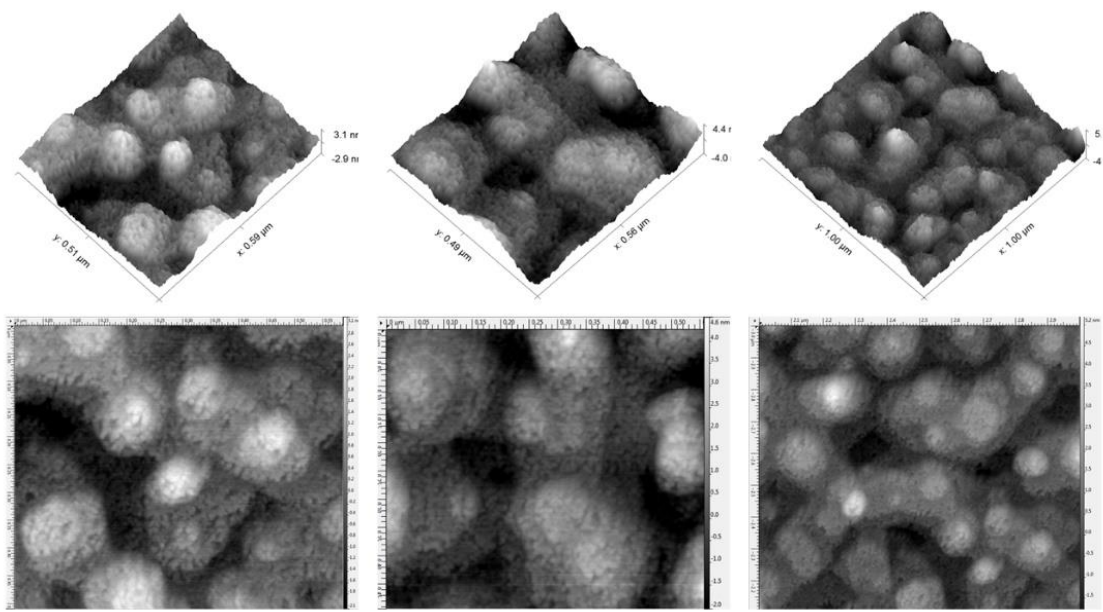

*Figure S8: AFM images of the epitaxial Sb$_2$Te$_3$ films used in this work. The surface is characterized by small crystalline grains with a diameter of about $15-20\, nm$ organized in agglomerates.*

## Bibliography

1. Hikami, S., Larkin, A. I. & Nagaoka, Y. Spin-Orbit Interaction and Magnetoresistance in the Two